\documentstyle[twocolumn,aps,psfig]{revtex}

\begin{document}
\draft
\flushbottom
\twocolumn[
\hsize\textwidth\columnwidth\hsize\csname @twocolumnfalse\endcsname

\title{ Optical second harmonic generation near a black hole horizon
as possible source of experimental information on quantum gravitational effects}
\author{Igor I. Smolyaninov }
\address{ Electrical and Computer Engineering Department \\
University of Maryland, College Park,\\
MD 20742}

\date{\today}
\maketitle
\tightenlines
\widetext
\advance\leftskip by 57pt
\advance\rightskip by 57pt

\begin{abstract}
Optical second harmonic generation near a black hole horizon is suggested as a source of experimental information on quantum gravitational effects. While absent in the framework of general relativity, second harmonic generation appears in the toy models of sonic and electromagnetic black holes, where spatial dispersion at high frequencies for waves boosted towards the horizon is introduced. Localization effects in the light scattering from random fluctuations of matter fields and space-time metric near the black hole horizon produce a pronounced peak in the angular distribution of second harmonics of light in the direction normal to the horizon. Such second harmonic light has the best chances to escape the vicinity of the black hole. This phenomenon is similar to the well-known strong enhancement of diffuse second harmonic emission from a randomly rough metal surface in the direction normal to the surface. 
\end{abstract}

]
\narrowtext

\tightenlines

Wave propagation and localization phenomena in random media have been the topic of extensive studies during the last years \cite{1}. One of the most striking examples of such phenomena is the strong and narrow peak of diffuse second harmonic light emission observed in the direction normal to a randomly rough metal surface (see Fig.1(a)). This peak is observed under the coherent illumination at any angle. This effect was initially predicted theoretically \cite{2} and later observed in the experiment \cite{3}. The enhanced second harmonic peak normal to the mean surface arises from the fact that a state of momentum {\bf k} introduced into a weakly localized system will encounter a significant amount of backscattering into states of momentum centered about {\bf -k}. When these surface {\bf k} and {\bf -k} modes of frequency $\omega $ interact through an optical nonlinearity to generate $2\omega $ radiative modes, the $2\omega $ light has nonzero wave vector components only perpendicular to the mean surface. The angular width of the normal peak can be as small as a few degrees, and its amplitude far exceeds the diffuse omnidirectional second harmonic background.   

Very recently it was realized \cite{4} that many results obtained in the optics of random media may be applicable to the case of light propagation in stochastic space-time metrics. This is possible because of an analogy between the propagation of light in matter and in curved space-time. It is well known that Maxwell equations in a general curved space-time background $g_{ik}(x,t)$ are equivalent to the phenomenological Maxwell equations in the presence of matter background with nontrivial electric and magnetic permeability tensors 
$\epsilon _{ij}(x,t)$ and $\mu _{ij}(x,t)$ \cite{5}. In this analogy, the event horizon corresponds to a surface of singular electric and magnetic permeabilities. There are quite a few papers which consider solid-state toy models of electromagnetic \cite{6} and sonic \cite{7} black holes. In the absence of established quantum gravitation theory such toy models are helpful
in understanding electromagnetic phenomena in curved space-time, such as Hawking radiation \cite{8} and Unruh effect \cite{9}. Introducing a natural high-frequency cutoff via modified dispersion relation for light or sound waves in the respective media, these papers demonstrated that the Hawking radiation does not depend on the arbitrary high frequency behavior of the theory. 

In this paper I am going to consider localization effects in light interaction with a black hole horizon. I am going to show that introduction of spatial dispersion at high frequencies leads to the appearance of surface electromagnetic modes near the horizon. Such modes are absent in the classical general relativity framework (there are no zero-geodesics along the surface of the black hole in the Schwarzschild metric). In combination with expected quantum fluctuations of space-time metric near the horizon, existence of surface electromagnetic modes allows to draw an analogy between localization effects in light interaction with the black hole horizon and with the rough metal surface (Fig.1). In this analogy quantum fluctuations of space-time metric (which are equivalent to fluctuating electric and magnetic permeability tensors $\epsilon _{ij}(x,t)$ and $\mu _{ij}(x,t)$) play the role of surface roughness. Weak localization effects in the scattering of surface optical modes should produce a pronounced peak in the angular distribution of second harmonics of light in the direction normal to the horizon. Because of the propagation direction perpendicular to the surface, such second harmonic light has the best chances to escape the vicinity of the black hole. This phenomenon is similar to the observed strong enhancement of second harmonic emission from a randomly rough metal surface in the direction normal to the surface. Since no such effect may be expected in the case of classical general relativity, experimental observation of optical second harmonic generation near a black hole horizon could be a unique source of experimental information on quantum gravity. 

To begin with, let us recall the formal analogy between the Maxwell equations in the Schwarzschild space-time background $g_{ik}$ and the phenomenological Maxwell equations in the presence of matter background with some electric and magnetic permeability tensors $\epsilon _{ij}$ and $\mu _{ij}$ (see for example \cite{4}). The metric of the Schwarzschild space-time is given by

\begin{equation} 
ds^2=(1-\frac{2M}{r})dt^2-(1-\frac{2M}{r})^{-1}dr^2-r^2(d\theta ^2+sin^2\theta d\phi ^2), 
\end{equation}

where we adopt units such that $k=c=1$. If we define $r=\rho +M+M^2/4\rho $, we can rewrite the metric as

\begin{equation} 
ds^2=f_1(\rho )dt^2-f_2^2(\rho )(d\rho ^2+\rho ^2d\theta ^2+\rho ^2sin^2\theta d\phi ^2), 
\end{equation}

\begin{figure}[tbp]
\centerline{
\psfig{figure=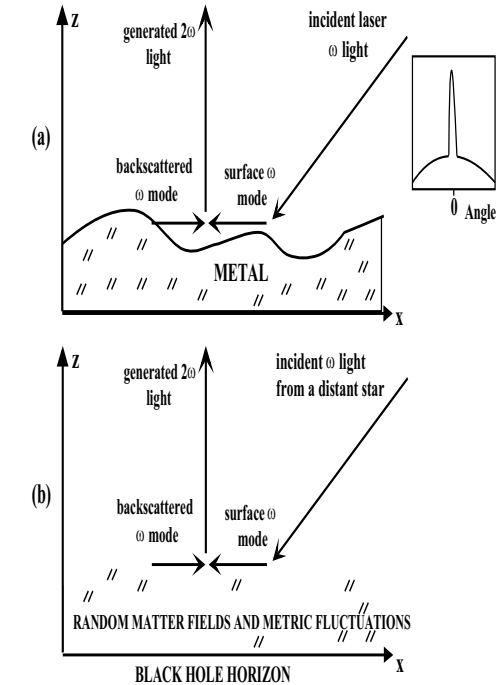,width=9.0cm,height=12.0cm,clip=}
}
\caption{ (a) Strong enhancement of second harmonic emission from a randomly rough metal surface in the direction normal to the surface. A typical angular distribution of the diffuse second harmonic light is shown in the inset. (b) Second harmonic generation near the black hole horizon. Light from a distant star provides the source of coherent illumination. 
}

\label{fig1}
\end{figure}

where $f_1(\rho )=1-2M/r$ and $f_2(\rho )=r/\rho $. We can bring this metric into a Cartesian form using a coordinate transformation

\begin{equation} 
x^1=\rho sin\theta cos\phi ,   
\end{equation}

\begin{equation}
x^2=\rho sin\theta sin\phi , 
\end{equation}

\begin{equation}
x^3=\rho cos\theta, 
\end{equation}

so it looks like

\begin{equation} 
ds^2=f_1(\rho )dt^2-f_2^2(\rho )((dx^1)^2+(dx^2)^2+(dx^3)^2) 
\end{equation}

If we consider this space-time acting as a medium, the corresponding effective electric and magnetic permeability tensors are expressed as

\begin{equation} 
\epsilon _{ik}=\mu _{ik}=\frac{f_2(\rho )}{f_1^{1/2}(\rho )}\delta _{ik}
\end{equation}

We are interested in physics at small distances z from the black hole horizon, so we can introduce coordinate transformation $\rho =M/2+z$ and get the effective Rindler geometry \cite{6} in the vicinity of the horizon

\begin{equation} 
ds^2=\alpha ^2z^2dt^2-dz^2-dx^2-dy^2, 
\end{equation}

where $M=(4\alpha )^{-1}$. The effective electric and magnetic permeability tensors will now look as

\begin{equation} 
\epsilon _{ik}=\mu _{ik}=\frac{1}{\alpha z}\delta _{ik}
\end{equation}

Wave equation in such an effective medium looks the same for both H and E polarization states of light \cite{10}:

\begin{equation} 
\Delta \vec{H}+\frac{\epsilon \mu \omega ^2}{c^2}\vec{H}+\frac{1}{\epsilon }(\nabla \epsilon \cdot rot\vec{H})=0
\end{equation}

Let us try to search for solutions which on a small scale look as if they 
propagate along x-coordinate ($\sim e^{i\kappa x}$). Introducing new variable 
$u = \epsilon ^{-1/2}H$, we can rewrite the wave equation (10) as

\begin{equation} 
\frac{\partial ^2u}{\partial z^2}+(\frac{1}{2\epsilon }\frac{\partial ^2\epsilon }{\partial z^2}-\frac{3}{4\epsilon ^2}(\frac{\partial \epsilon }{\partial z})^2+\frac{\epsilon \mu \omega ^2}{c^2}-\kappa ^2)u=0
\end{equation}

After substitution $\epsilon =\mu =1/\alpha z$, equation (11) takes the form similar to the one-dimensional Schrodinger equation for a $1/z^2$ potential well:

\begin{equation} 
\frac{\partial ^2u}{\partial z^2}+((\frac{1}{4}+\frac{\omega ^2}{\alpha ^2c^2})z^{-2}-\kappa ^2)u=0
\end{equation}

In the limit $z\rightarrow 0$ its solutions behave as

\begin{equation} 
H=H_0 cos(\frac{\omega }{\alpha c}ln(z/z_0)), 
\end{equation}

Thus, the z-component of the spatial frequency diverges near the horizon, and all the wave functions have infinite number of zeros near $z=0$ (there is no ground state). We have obtained a well-known result that there is no zero-geodesics along the surface of the black hole in the Schwarzschild metric.

On the other hand, as is well known from the general properties of Schrodinger equation \cite{11}, the case of $1/z^2$ potential is a boundary separating potential wells for which the finite ground state may or may not exist. The ground state exists for any attractive potential that is weaker then $1/z^2$. Thus, introduction of any cutoff for whatever reason in the divergent $\sim 1/z$ behavior of the effective electric $\epsilon $ and magnetic $\mu $ permeability tensors near the black hole horizon leads to appearance of a ground state among solutions of equation (11). This ground state may be considered as a "guided surface mode" in the effective surface waveguide near the horizon. 

Quite a few theoretical models introduce such cutoffs. 
Change in the character of sound wave trajectories near the horizon due to the high-frequency cutoff via modified dispersion relation in the toy model of sonic black hole has been reported by Unruh \cite{7}. His numerical simulations of sound wave trajectories near the horizon indicate appearance of new type of trajectories, which do not fall into the event horizon, but eventually reflect back. Reznik \cite{6} has reported similar simulations in the toy model of dielectric electromagnetic black hole. He conjectured that introduction of the high- frequency cutoff via modified dispersion relation is equivalent to a "frequency-dependent" geometry, where the effective geometry experienced by the particle depends on its energy. This point of view is natural within the scope of dielectric toy models, since electric and magnetic permeability tensors of all materials are frequency-dependent. Possible effects of frequency-dependent geometries are being considered in the current literature (see for example \cite{12}) as possible manifestations of extra spatial dimensions.

Another example of modified metric (and, hence, effective $\epsilon $ and $\mu $) may be found in \cite{13}, where a new kind of static, spherically symmetric solution to Einstein's equations has been introduced. In this solution the horizon is considered as a critical surface of a quantum phase transition, while the interior and exterior regions far from the horizon are locally flat. In this metric the equivalent $\epsilon $ and $\mu $ distributions form an effective surface waveguide.

We should also remember that equations (10-13) are derived from the linear Maxwell equations, which do not take into account electron-positron pair creation by strongly blue-shifted photons interacting with matter fields and metric fluctuations near the horizon. Exact analysis of nonlinear Maxwell equations near the horizon would also require taking into account unknown contributions to $\epsilon $ and $\mu $ from the random matter fields possibly absorbed by the black hole in the past. It is sufficient for our present analysis to establish the presence of a "guided modes" near the horizon surface, and the fact of strong nonlinear interaction of these modes with each other (due to the pair creation and annihilation near the horizon). These modes also experience scattering by the random matter fields (and possibly metric fluctuations), so there is close resemblance of this situation with the situation of light interacting with randomly rough metal surface, where surface polaritons play the part of the guided modes.

Subsequent consideration of localization effects may closely replicate the derivation given by McGurn et al. \cite{2} for a randomly rough surface. Let us consider the case of a black hole that presently does not absorb much of an incoming light and matter from its immediate vicinity, so that the optical nonlinearity associated with the horizon is confined to a length scale smaller than the wavelength of light in vacuum. At the same time, let us assume some random matter distribution near the horizon and the presence of hitherto unknown quantum metric fluctuations. In such a case, the optical nonlinearity of the surface may be introduced as a set of boundary conditions on the $2\omega $ fields at the random interface, similar to the case of rough metal-vacuum interface, since in both cases the nonlinearity near the surface arises from rapid variations in the electromagnetic fields and charge densities, and the breaking of the translational symmetry of the system by the surface. 

Outgoing second harmonic field at large z can be written in the form

\begin{equation} 
H_2=\int \frac{d\kappa }{2\pi }S(\kappa )e^{i\kappa x+iqz} , 
\end{equation}

where $q=(\frac{4\omega ^2}{c^2}-\kappa ^2)^{1/2}$. Similar to \cite{2}, expression for the angular $2\omega $ field distribution $S(\kappa )$ in terms of the $\omega $ fields can be obtained by considering the boundary conditions on the $2\omega $ fields due to the nonlinear polarization near the horizon. Assuming that nonlinear polarization in the surface region depends only on z, the following boundary conditions may be obtained by integration of nonlinear Maxwell equations \cite{2}:

\begin{equation} 
E_x(2\omega )\approx -\mu _1\frac{\partial }{\partial x}(E_z(\omega ))^2-\mu _ 2\frac{\partial }{\partial x}(E_x(\omega ))^2, 
\end{equation}

\begin{equation} 
H_y(2\omega )\approx -\frac{2i\omega }{c}\mu _3E_x^2(\omega ), 
\end{equation}

where $\mu _1$, $\mu _2$, and $\mu _3$ are the functions of $\epsilon (z,\omega )$, $\epsilon (z,2\omega )$, and the random charge and current density distributions near the surface. Detailed expressions for $\mu _1$, $\mu _2$, and $\mu _3$ can be found in \cite{2}. A very important result of \cite{2} is that  
the angular distribution of diffusely generated second harmonic of light from randomly rough surface is not sensitive to the ratios $\mu _1/\mu _3$ and $\mu _1/\mu _2$ (which are not known, and very difficult to obtain experimentally). This fact has to do with localization effects in scattering of $\omega $ and $2\omega $ fields. The localization of $\omega $ surface modes contributes a peak in the angular distribution of the intensity of generated second harmonic centered about the normal to the mean surface. In the scattering process which gives rise to this peak the incident $\omega $ fields excite weakly localized $\omega $ surface modes. Interaction of this modes with the backscattered surface modes creates $2\omega $ light that propagates in the direction normal to the surface. This happens because the total tangential component of the momentum of the surface modes is conserved. Another peak in the angular distribution of outgoing $2\omega $ fields is directed opposite to the motion of the original $\omega $ incident light. This peak is due to the weak localization of $2\omega $ surface modes \cite{2}. Variations in the $\mu _1/\mu _3$ and $\mu _1/\mu _2$ ratios cause some variations in the relative intensities of these two peaks and a weak omnidirectional diffuse second harmonic background, but the general shape of the angular distribution remains basically the same. Theoretical findings of \cite{2} were successfully confirmed in the experiment \cite{3}, which indicates that the described effect is a general property of weakly disordered systems that support surface optical modes. 

Another important analogy between the solid state situation in \cite{2,3} and light interaction with an isolated black hole is that distant stars provide the source of coherent optical illumination, which is to some extent similar to the laser illumination used in the solid state experiments \cite{3}.
We should also emphasize that second harmonic light propagating perpendicular to the horizon has the best chances to escape the vicinity of a black hole, thus becoming the dominant part of its visible emission. This circumstance should also reduce the omnidirectional second harmonic background compared to the intensity of the second harmonic peak in the normal direction. 

The described optical second generation may be used to obtain unique experimental information on quantum gravitational effects.
It may also be used to detect an individual black hole that does not have close neighbors providing it with a constant inflow of matter. It could reveal itself as a weak light source that has an anomalously strong blue shift (which is a more "astronomical" way of describing second harmonic light). The total brightness of the source should be proportional to the surface area of the black hole and the square of the total incident light intensity provided by nearby stars. The second harmonic conversion efficiency should depend on how close the incident $\omega $ light can get to the horizon before been scattered. In the best case scenario it can be as good as the second harmonic conversion efficiency in solid state physics situations, where it can reach up to a few percents. These arguments provide reasonably optimistic prognosis for the detection of second harmonic light generated by black holes in the dense stellar associations. 

I should also mention recent astronomical observations of second harmonic of cyclotron line in the spectra of some high mass bright X-ray sources, such as 4U1907+09 obtained by the BeppoSAX satellite \cite{14}. In the spectrum of this source the cyclotron line at 19 keV is accompanied by the second harmonic line at 39 keV. It is interesting that not every bright X-ray source has such a feature in its spectrum. For example, X-ray pulsar 4U 1538-52 observed by the same satellite exhibit only the cyclotron line at 21 keV with no second harmonic present \cite{15}. It is widely believed that many bright X-ray sources are powered by black holes, so these observations may be quite relevant for the theoretical findings described above.  

In conclusion, optical second harmonic generation near a black hole horizon is suggested as a source of experimental information on quantum gravitational effects. While absent in the framework of general relativity, second harmonic generation appears in the toy models of sonic and electromagnetic black holes, where spatial dispersion at high frequencies for waves boosted towards the horizon is introduced. Localization effects in the light scattering from random fluctuations of matter fields and space-time metric near the black hole horizon produce a pronounced peak in the angular distribution of second harmonics of light in the direction normal to the horizon. Such second harmonic light has the best chances to escape the vicinity of the black hole. This phenomenon is similar to the well-known strong enhancement of diffuse second harmonic emission from a randomly rough metal surface in the direction normal to the surface.

\end{document}